\documentclass[sigconf]{acmart}
\usepackage{listings}
\usepackage{color}
\usepackage{bbm}
\usepackage{subfigure}
\usepackage{graphicx}
\usepackage{float}
\usepackage{enumitem}
\usepackage{threeparttable}
\usepackage{ulem}

\makeatletter
\renewcommand{\@thesubfigure}{\hskip\subfiglabelskip}
\makeatother
\usepackage{multirow}
\usepackage[linesnumbered, ruled, vlined]{algorithm2e}
\definecolor{dkgreen}{rgb}{0,0.6,0}
\definecolor{gray}{rgb}{0.5,0.5,0.5}
\definecolor{mauve}{rgb}{0.58,0,0.82}

\AtBeginDocument{%
  \providecommand\BibTeX{{%
    \normalfont B\kern-0.5em{\scshape i\kern-0.25em b}\kern-0.8em\TeX}}}

\copyrightyear{2023}
\acmYear{2023}
\setcopyright{acmlicensed}\acmConference[KDD '23]{Proceedings of the 29th ACM SIGKDD Conference on Knowledge Discovery and Data Mining}{August 6--10, 2023}{Long Beach, CA, USA}
\acmBooktitle{Proceedings of the 29th ACM SIGKDD Conference on Knowledge Discovery and Data Mining (KDD '23), August 6--10, 2023, Long Beach, CA, USA}
\acmPrice{15.00}
\acmDOI{10.1145/3580305.3599367}
\acmISBN{979-8-4007-0103-0/23/08}

\begin{document}

\title{GMOCAT: A Graph-Enhanced Multi-Objective Method for Computerized Adaptive Testing}

\author{Hangyu Wang}
\email{hangyuwang@sjtu.edu.cn}
\affiliation{%
  \institution{Shanghai Jiao Tong University}
  \country{Shanghai, China}
}

\author{Ting Long}
\email{longting@jlu.edu.cn}
\authornote{Work was done in Shanghai Jiao Tong University.}
\affiliation{%
  \institution{Jilin University}
  \country{Changchun, China}
}

\author{Liang Yin}
\email{yinla@apex.sjtu.edu.cn}
\affiliation{%
  \institution{Shanghai Jiao Tong University}
  \country{Shanghai, China}
}

\author{Weinan Zhang}
\authornote{Corresponding author.}
\email{wnzhang@sjtu.edu.cn}
\affiliation{%
  \institution{Shanghai Jiao Tong University}
  \country{Shanghai, China}
}

\author{Wei Xia}
\email{xiawei24@huawei.com}
\affiliation{%
  \institution{Huawei Noah’s Ark Lab}
  \country{Shenzhen, China}
}

\author{Qichen Hong}
\email{hongqichen@huawei.com}
\affiliation{%
  \institution{Huawei CBG Edu AI Lab}
  \country{Shenzhen, China}
}

\author{Dingyin Xia}
\email{xiadingyin@huawei.com}
\affiliation{%
  \institution{Huawei CBG Edu AI Lab}
  \country{Shenzhen, China}
}

\author{Ruiming Tang}
\email{tangruiming@huawei.com}
\affiliation{%
  \institution{Huawei Noah’s Ark Lab}
  \country{Shenzhen, China}
}

\author{Yong Yu}
\authornotemark[2]
\email{yyu@apex.sjtu.edu.cn}
\affiliation{%
  \institution{Shanghai Jiao Tong University}
  \country{Shanghai, China}
}


\renewcommand{\shortauthors}{Hangyu Wang, et al.}

\begin{abstract}
Computerized Adaptive Testing (CAT) refers to an online system that adaptively selects the best-suited question for students with various abilities based on their historical response records. Compared with traditional CAT methods based on heuristic rules, recent data-driven CAT methods obtain higher performance by learning from large-scale datasets. However, most CAT methods only focus on the quality objective of predicting the student ability accurately, but neglect concept diversity or question exposure control, which are important considerations in ensuring the performance and validity of CAT. Besides, the students' response records contain valuable relational information between questions and knowledge concepts. The previous methods ignore this relational information, resulting in the selection of sub-optimal test questions. To address these challenges, we propose a \textbf{G}raph-Enhanced \textbf{M}ulti-\textbf{O}bjective method for \textbf{CAT} (\textbf{GMOCAT}). Firstly, three objectives, namely quality, diversity and novelty, are introduced into the Scalarized Multi-Objective Reinforcement Learning framework of CAT, which respectively correspond to improving the prediction accuracy, increasing the concept diversity and reducing the question exposure. We use an Actor-Critic Recommender to select questions and optimize three objectives simultaneously by the scalarization function. Secondly, we utilize the graph neural network to learn relation-aware embeddings of questions and concepts. These embeddings are able to aggregate neighborhood information in the relation graphs between questions and concepts. We conduct experiments on three real-world educational datasets. The experimental results show that GMOCAT not only outperforms the state-of-the-art methods in the ability prediction, but also achieve superior performance in improving the concept diversity and alleviating the question exposure. Our code is available at \url{https://github.com/justarter/GMOCAT}.
\end{abstract}


\maketitle

\vspace{-5pt}
\section{Introduction}
\vspace{-2pt}
With the rapid development of Internet technology, Computerized Adaptive Testing (CAT) gradually releases the repetitive work with paper-and-pencil tests \cite{appofcat}. CAT is an online test that can accurately measure the student ability by continuously feeding the most suitable questions to students \cite{cat:primer}. CAT has been applied in many large-scale educational examination scenarios, e.g., GMAT \cite{gmat} and GRE \cite{gre}, to increase student engagements \cite{cat:overview}.

\begin{figure}
    \subfigure[(a)]{
    \centering
    \includegraphics[width=.45\linewidth]{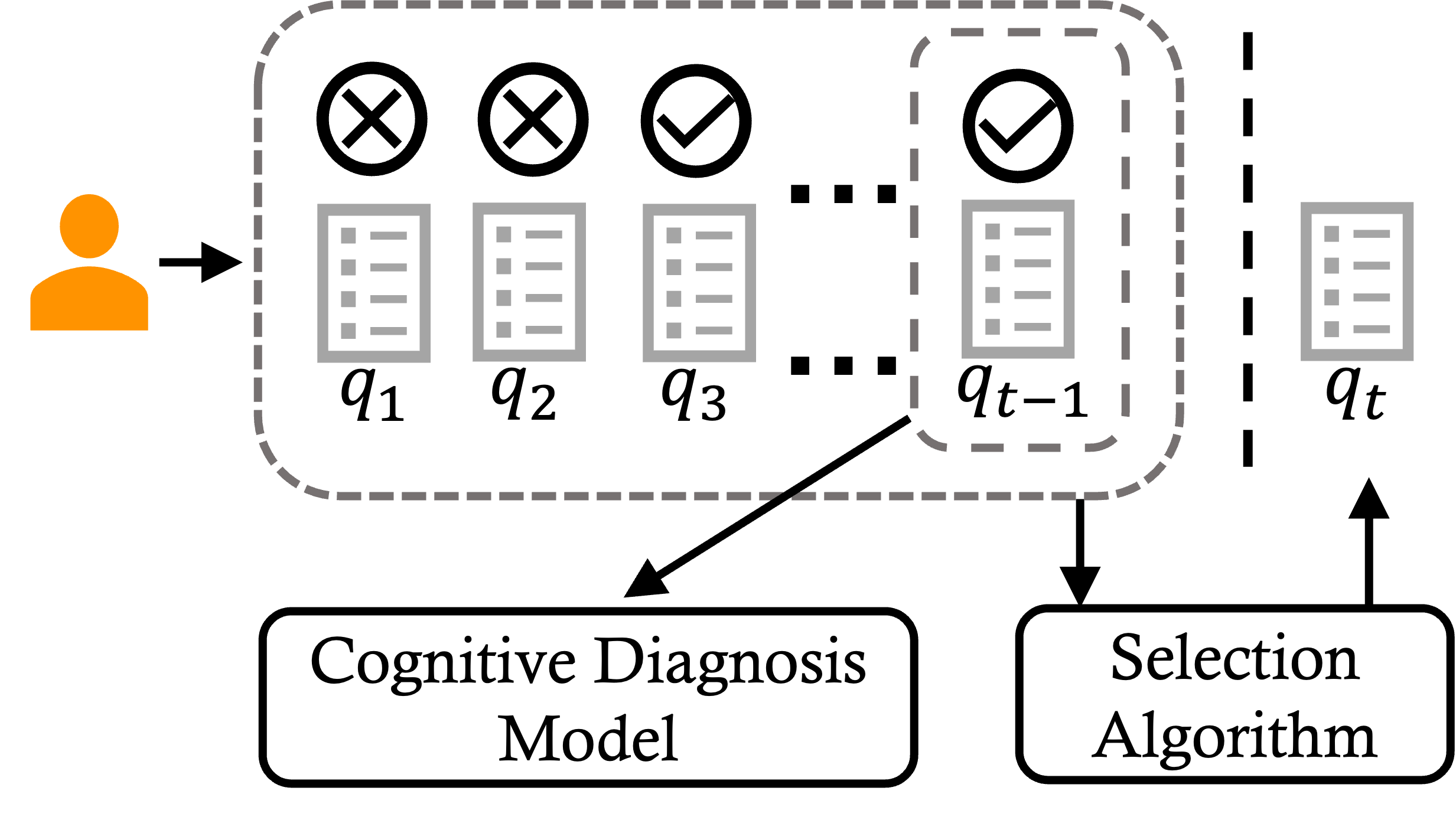}
    \label{fig:catprocess}
    }\quad
    \subfigure[(b)]{
    \centering
    \includegraphics[width=.45\linewidth]{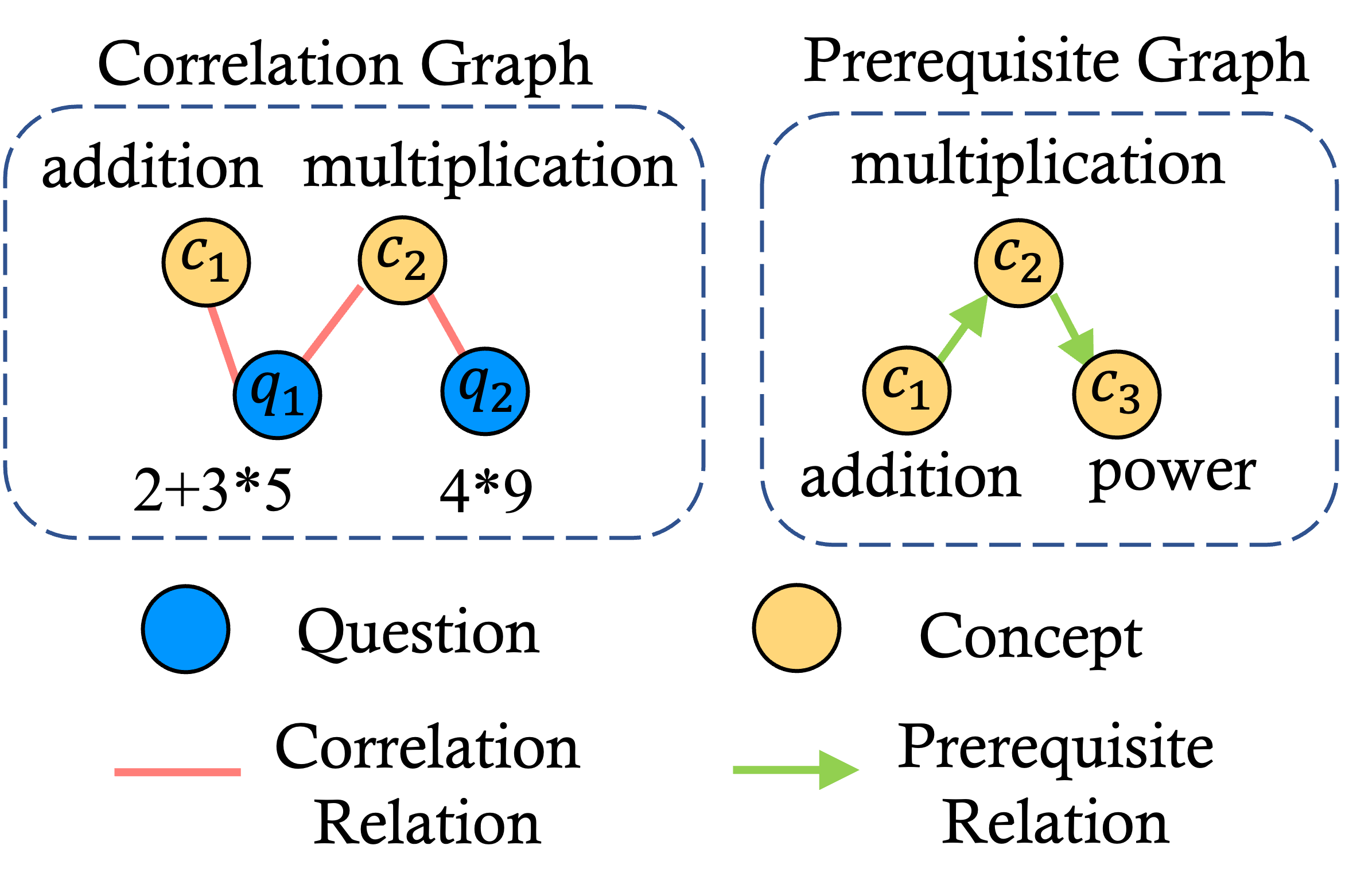}
    \label{fig:graphexample}
    }
    \vspace{-10pt}
    \caption{(a) The workflow of CAT: at step $t$, the selection algorithm selects the next question $q_{t}$ based on the historical response records. (b) The toy examples of correlation and prerequisite graphs.}
    \vspace{-15pt}
\end{figure}

Figure \ref{fig:catprocess} shows an example of the CAT procedure. A CAT system usually consists of two main components, which work iteratively: (1) \textbf{Cognitive Diagnosis Model (CDM)}, which captures a student's ability using her responses to questions \cite{cdm}. The simplest CDM is Item Response Theory (IRT) \cite{irt}, using an item response function to approximate the student's real ability. Deep learning-based CDMs, such as NeuralCDM (NCD), apply neural networks to model interactions between students and questions \cite{ncd}. (2) \textbf{Selection Algorithm}, which selects the most suitable question for a student based on her historical response records. Traditional static algorithms usually use heuristic rules to select questions with the largest information \cite{appofirt} or with the largest expected model change \cite{maat}. These algorithms are usually greedy for one step but lack a long-term perspective. In recent years, data-driven approaches that learn selection rules from large-scale datasets have also emerged \cite{bobcat}. The selection algorithm helps CDM evaluate the student ability more efficiently by selecting the best-suited questions.

As a question selector, the selection algorithm plays a crucial role in the above CAT process, thus we focus on designing an effective data-driven selection algorithm in this paper. In recent years, data-driven selection algorithms have been proposed from the perspectives of meta learning \cite{bobcat} or Reinforcement Learning (RL) \cite{ncat}. However, these studies only focus on the quality objective of predicting the student ability, which is insufficient in real-world scenarios \cite{cat:overview}. We argue that the single-objective method suffers from two main limitations: (1) a lack of concept diversity. A good examination evaluates students' abilities on related but diverse knowledge concepts \cite{maat}. For example, at the end of each semester, the final exam for mathematics usually covers concepts in algebra, geometry, etc. Unfortunately, previous algorithms are suboptimal due to the neglect of diversity issues, leading to very limited concepts. (2)a lack of novelty. CAT keeps reusing all questions in the question pool for different students, causing some questions to be selected too frequently. The overexposure of test questions will reduce their novelty and change the student's test-taking behavior \cite{han2018components}. For example, overexposed questions can be known to many students, which could inflate the scores of subsequent students. Horribly, \citet{bobcat} found that most selection algorithms prefer a part of questions, resulting in an excessive exposure rate. That is not practical in real-world CAT systems. In conclusion, quality, diversity and novelty are all important and deserve attention in CAT. Although some of previous studies \cite{maat,Randomesque} have noticed these problems, they only focus on parts of them, and none have addressed these problems from a unified perspective.


To address the above shortcomings, we propose a \textbf{G}raph-Enhanced \textbf{M}ulti-\textbf{O}bjective method for \textbf{CAT} (\textbf{GMOCAT}). Firstly, we formalize the CAT procedure as a Multi-Objective Markov decision process (MOMDP) and then introduce a Scalarized Multi-Objective Reinforcement Learning (Scalarized MORL) framework into the CAT setting. 
Compared with the greedy methods for CAT, the RL framework has been proven to explore more appropriate questions for students from a long-term view \cite{ncat}.

In a unified framework, our GMOCAT considers the following three objectives: i) \textbf{Quality} predicts the student ability accurately. ii) \textbf{Diversity} diversifies knowledge concepts in recommended questions. iii) \textbf{Novelty} controls the question exposure. In view of these three objectives, we design three rewards, namely, quality, diversity and novelty rewards. We design an Actor-Critic Recommender to select questions, which optimizes three objectives simultaneously with the scalarization function. A naive and widely used approach in MORL is to simply modify the environment to return a scalar weighted reward and optimize the policy by a single-objective method \cite{mooinlearning}. In contrast to this, we improve the single-objective algorithm by extending the value and reward function to be vectorized. With the more fine-grained and vectorized feedback, the Actor-Critic Recommender doesn't confuse the objectives and achieves better performance \cite{cantintoonereward, preguidemorl}.



To further improve the effectiveness of CAT, the relational information between the questions and knowledge concepts can be utilized, since this information is closely related to the objectives in CAT. For example, the diversity objective requires to select questions that contain a variety of knowledge concepts. In our work, we mainly consider two types of relations: correlation and prerequisite, which are shown in Figure \ref{fig:graphexample}. Correlation relation exists between a question and its related concepts, and prerequisite relation involves a pair of concepts, implying that one concept should be learned logically before the other (e.g. \textit{multiplication} is the successor of \textit{addition}). This relational information is crucial for selecting appropriate questions, but has been overlooked in previous work. Therefore, we employ relational information for question selection in our framework. In particular, we use graph attention networks to extract and aggregate neighborhood information from the multiple relation graphs.



In summary, our key contributions are listed as follows:
\begin{itemize}[leftmargin=2pt]
    \item We consider three important objectives in CAT: Quality, Diversity, and Novelty, and integrate them into a unified MORL framework. We also propose three rewards to quantify the feedback from three objectives. To our knowledge, this is the first work to apply MORL in CAT.
    \item We introduce relation graphs into CAT and learn relation-aware embeddings to help select more appropriate questions. It's also the first attempt to use relation graphs to aid in question selection.
    \item We conduct extensive experiments on three real-world educational datasets. The experimental results show that our method achieves a more accurate ability estimate than the state-of-the-art methods. Meanwhile, our proposed approach also significantly improves the concept diversity and reduces the question exposure.
\end{itemize}

\vspace{-10pt}
\section{Related Work}
\subsection{Computerized Adaptive Testing}
Computerized Adaptive Testing (CAT) has two main components: a Cognitive Diagnosis Model (CDM) and a Selection Algorithm. In traditional CAT systems, a widely used CDM is Item Response Theory (IRT) \cite{irt}, which estimates the student ability by predicting her response to questions. The recently emerged Neural Cognitive Diagnosis Model (NCD) utilizes the neural network to model the student-question interactions \cite{ncd}.


This paper focuses on the selection algorithm. The most widely used algorithm utilizes Maximum Fisher Information (MFI) \cite{appofirt} to select questions. Alternatively, Kullback-Leibler Information (KLI) \cite{kli} calculates the integral over an ability interval to pick questions. These heuristic algorithms are designed for specific CDMs, such as IRT. To alleviate this problem, \citet{maat} proposed a model-agnostic algorithm, MAAT, that leverages active learning for question selection. They also design an extra module to enforce concept diversity. RAT \cite{rat} benefits the selection algorithm by capturing multiple aspects of the student ability. After that, more deep-learning based and data-driven algorithms have been developed. For example, BOBCAT \cite{bobcat} is a meta learning-based method that couples CDM and selection algorithm together in a bilevel optimization problem. NCAT \cite{ncat} is a reinforcement learning-based method that utilizes an attention-based DQN to select questions. NCAT also controls the question exposure by sampling from the Boltzmann distribution \cite{boltzmanndistribution}. The above methods only consider either the importance of diversity or novelty without the combination of these two parts. To the best of our knowledge, few existing works have well established the multi-objective framework for CAT. 

\vspace{-5pt}
\subsection{Multi-Objective Optimization}
Multi-Objective Optimization aims to reach Pareto Optimality while optimizing multiple objectives simultaneously \cite{moogenetic}. Multi-objective problems can be solved by various methods, such as genetic algorithms \cite{moogenetic}, evolutionary algorithms \cite{mooincat} or Multi-Objective RL algorithms \cite{modrl}. In CAT, \citet{mooincat} proposed optimizing test length and accuracy by a multi-objective evolutionary algorithm. However, this method has not been verified on a real-world dataset. As far as we know, the most similar method to ours is DRE \cite{mooinlearning} in the field of adaptive learning, which integrates three rewards into one, and uses a DQN strategy. In contrast, we apply Scalarized Multi-Objective policy gradient method to maintain mutual independence of objectives.

\vspace{-5pt}
\subsection{Knowledge Graph}
Knowledge Graph contains a large amount of information with nodes (entities, e.g. questions or concepts) and edges (relations, e.g. prerequisite) \cite{reviewofrelationgraph}. The relation graph, as a type of knowledge graph, has been used in many fields with various graph representation learning \cite{rcd,gkt,cseal}. For example, GKT \cite{gkt} uses Graph Neural Network (GNN) \cite{gnn} with a graph-like knowledge structure for knowledge tracing. RCD \cite{rcd} uses Graph Attention Network (GAT) \cite{gat} to aggregate multi-level information for cognitive diagnosis and CSEAL \cite{cseal} designs a graph-based cognitive navigation for adaptive learning. To our best knowledge, we are the first to involve the relation graph in the CAT setting. 

\vspace{-5pt}
\section{Preliminaries}

\subsection{Terminologies} 
\definition \textbf{Response Record}. In CAT process, for the student $i$, her response record at test step $t$ is denoted as ($q^i_t$,$c^i_t$,$y^i_t$), where $q^i_t$ denotes the question responded by the student at step $t$, and $c^i_t$ denotes the concept covered by this question, and $y^i_t$ denotes the student's response. $y^i_t$ is 1 if the response is correct, and 0 otherwise. 

We define two types of graphs to represent relations among questions and concepts.
Taking Figure \ref{fig:graphexample} as an example, we define the correlation graph and the prerequisite graph\footnote{If a dataset does not explicitly contain the graph, we can construct one with the method from Appendix \ref{app:graphconstruction}.}:

\vspace{-5pt}
\definition \textbf{Correlation Graph ($\mathcal{G}_{qc}$)}: is an undirected bipartite graph to represent the relations between the questions and their related knowledge concepts. The set of nodes in $\mathcal{G}_{qc}$ is composed by questions and concepts. An arbitrary edge in the edge set connects a question and one of its related concepts. 
\vspace{-5pt}
\definition \textbf{Prerequisite Graph ($\mathcal{G}_{cc}$)}: is a directional graph to represent the prerequisite relations between concepts. The set of nodes in $\mathcal{G}_{cc}$ is composed of knowledge concepts, and an arbitrary edge in the edge set connects a concept and its prerequisite concept.



\vspace{-5pt}
\subsection{The CAT Process and Problem Setting}

CAT is composed of CDM and the selection algorithm, in which the former aims to estimate the student ability and the latter aims to select the best-suited question. Data-driven selection algorithms have been demonstrated to be superior to hand-designed ones \cite{bobcat}. Therefore, we focus on designing an effective data-driven selection algorithm in this paper.

The CAT process with the data-driven selection algorithm consists of training and testing phases. To train/test the selection algorithm, a regular setting \cite{ncat} is splitting the samples which contain student $i$'s test records into \textbf{Candidate Question Set} $\mathcal{D}^i_c$ and \textbf{Meta Question Set} $\mathcal{D}^i_m$, as illustrated in Appendix \ref{app:training/testing_phase}. 
Below, we take the single-objective selection algorithm as an example to analyze the training/testing phase in CAT process:


\textbf{Training Phase}. For each student $i$ in the training set, (1) at test step $t\in [1,T]$, the selection algorithm selects $q^i_t$ from candidate question set $\mathcal{D}^i_c$ based on her historical response records; (2) the student gives her response $y^i_t$ and CDM updates the current ability estimate $\theta^i_t$; (3) use $\theta^i_t$ and meta question set $\mathcal{D}^i_m$ to calculate a feedback(e.g., reward), which measures the accuracy of the ability estimate; (4) after $T$ iterations of the above process, we train the selection algorithm to maximize the feedback.

\textbf{Testing Phase\footnote{Non-data-driven selection algorithms do not require training and can be tested directly}}. For a new student $j$ in the testing set, stages (1) and (2) are the same as the training phase. Stage (3) is to evaluate the accuracy of $\theta^j_t$ on her meta question set $\mathcal{D}^j_m$. The selection algorithm is not trained during the testing phase.


In the CAT process, the student's real ability ${\theta}^i_{*}$ is assumed to remain constant \cite{psybehindcat}. From the above phases, we can find that the goal of single-objective method, so-called the quality objective, is to make the final estimate $\theta^i_T$ as close to real ability ${\theta}^i_*$ as possible by maximizing feedback. In our work, we not only consider the above quality objective, but also the diversity and novelty objectives, so as to optimize and evaluate CAT in multiple aspects. Overall, our target in this paper is \textit{the design of selection algorithm, which selects the best-suited question at each step to achieve three objectives}.

\vspace{-5pt}
\section{Method}

\begin{figure*}
    \centering
    \vspace{-8pt}
    \includegraphics[width=\linewidth]{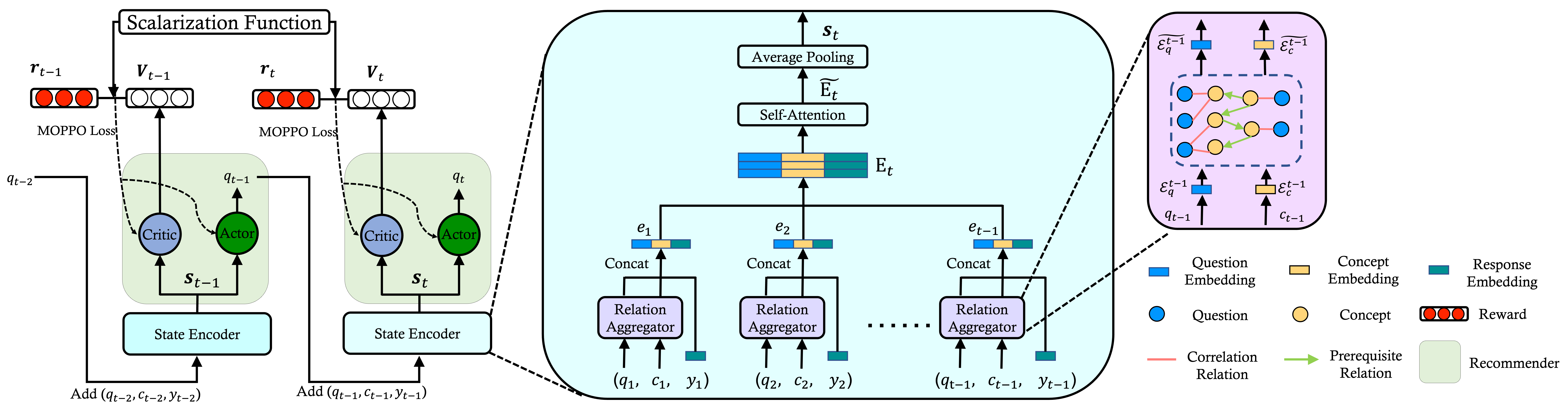}
    \caption{The overall framework of GMOCAT. Relation aggregator learns the relation-aware embeddings of questions and concepts. State encoder converts the historical response records into a low-dimensional state. Actor-critic recommender chooses the next question and then receives the multi-objective reward.}
    \vspace{-8pt}
    \label{fig:modelframework}
\end{figure*}

\subsection{Formulation and Overview}
As described in the Introduction, the CAT process is a complex and interdependent system . 
We model the CAT task as a sequential decision problem and formalize it as a Multi-Objective Markov decision process (MOMDP) \cite{modrl}. The RL framework can explore more suitable questions from a long-term view rather than the greedy approximation \cite{ncat,dai2021adversarial}. This MOMDP can be defined by tuples of $<\mathcal{S}, \mathcal{A}, \mathcal{P}, \mathbf{R}, \gamma>$, in which:
\begin{itemize} [leftmargin=2pt]
    \item $\mathcal{S}$: denotes the set of states that are used by the selection algorithm to select questions. Given a test step $t$, the state can be defined as $\mathbf{s}^i_t=f_{se}(\{(q^i_1,c^i_1,y^i_1),\ldots,(q^i_{t-1},c^i_{t-1},y^i_{t-1})\}) \in \mathcal{S}$, where $f_{se}$ is a state encoder that will be discussed in Section \ref{sec:stateencoder}. The state encoder takes the student $i$'s historical response records $\{(q^i_1,c^i_1,y^i_1),\ldots,(q^i_{t-1},c^i_{t-1},y^i_{t-1})\}$ as input and outputs state $\mathbf{s}^i_t$.
    \item $\mathcal{A}$: is a finite set of actions. At a given step $t$, an action is taken from the action space (i.e., the candidate question set) to select the question $q^i_t$ by the selection algorithm $\pi$.
    \item $\mathcal{P}$: denotes the transition probability of reaching next state $\mathbf{s}^i_{t+1}$ after selecting question $q^i_t$ on state $\mathbf{s}^i_t$. i.e., $\mathcal{P}(\mathbf{s}^i_{t+1}|\mathbf{s}^i_t,q^i_t)$.
    \item $\mathbf{R}$: $\mathcal{S} \times \mathcal{A} \mapsto \mathbb{R}^m$, denotes the instant reward function for the selection algorithm to select $q^i_t$ on state $\mathbf{s}^i_t$. Different from previous work, our reward is vector-valued. In our setting, $\mathbf{r}(\mathbf{s}^i_t, q^i_t)= [r_{qua}, r_{div},r_{nov}]$, which refers to the quality, diversity and novelty reward respectively. This reward will be explained in Section \ref{sec:moreward}.
    \item $\gamma \in [0,1]$: is the discounted factor that trades off the immediate and future rewards. 
\end{itemize}

So far, we have reconstructed the CAT process from the perspective of MORL. 

\textbf{MORL Formulation of CAT Process}: Let $n$ denote the number of students. For student $i$, at test step $t$, the selection algorithm $\pi$ selects question $q^i_t$ from her candidate question set based on the state $\mathbf{s}^i_t$, i.e., $\pi(q^i_t|\mathbf{s}^i_t)$. Then, it pushes $q^i_t$ to the student $i$ and receives multi-objective reward $\textbf{r}(\mathbf{s}^i_t,q^i_t)$. Finally, our multi-objective goal is to maximize the weighted-sum return $\mathcal{J}$:
\vspace{-5pt}
\begin{align}
    \max_{\pi} \mathcal{J} &= \max_{\pi} \frac{1}{n} \sum_{i=1}^{n} \left[ \mathbf{w}^T \left( \sum_{t'=1}^{T}\gamma^{t'}\mathbf{r}\left(\mathbf{s}^i_{t'},q^i_{t'}\right)\right) \right] \\
   &=\max_{\pi} \mathbb{E}_{i \sim \pi} \left[ \mathbf{w}^T \left( \sum_{t'=1}^{T}\gamma^{t'}\mathbf{r}\left(\mathbf{s}^i_{t'},q^i_{t'}\right)\right) \right]
   \label{eq:objectivefunction}
\end{align}
$\mathbf{w}$ is the scalarization function, which can be regarded as a weight vector whose element represents the importance of each objective. 

Under this goal, our GMOCAT consists of four components: a multi-objective reward, a relation aggregator, a self-attentive state encoder, and an actor-critic recommender. As it is shown in Figure \ref{fig:modelframework}, the relation aggregator uses relation graphs to learn the relation-aware embeddings of questions and concepts. Then, state encoder uses these embeddings to encode students' historical response records, and generates the state of GMOCAT. Subsequently, this state is fed to the actor-critic recommender, which is instructed by the multi-objective reward, to select the questions. 

To adequately explain the details of GMOCAT, we will first go through the multi-objective reward, self-attentive state encoder and actor-critic recommender. Lastly, we discuss how to use the relation aggregator to learn the relation-aware embeddings. Please note that \textit{in the following sections, we omit superscript $i$ as we discuss how to select questions for a single student}.

\subsection{Multi-Objective Reward} 
\label{sec:moreward}

In this part, we discuss how to design our multi-objective reward, which plays an important role in learning the optimal selection algorithm. As mentioned before, previous works only focus on the quality objective. However, such single-objective methods cannot satisfy the needs of CAT in practice. In this paper, we incorporate three domain-specific objectives into the reward design, including quality, diversity and novelty, to support the adaptive question selection.


\subsubsection{\textbf{Quality}}

An excellent selection algorithm should select the best-suited question to predict the student's ability accurately. For an arbitrary student, since her real ability is unknown, we can use her meta question set to measure the error of ability estimate ${\theta}_t$. Specifically, at test step $t$, we use ${\theta}_t$ to calculate the prediction accuracy on her meta question set, which is denoted as $ACC({\theta}_t)$.
A higher value of $ACC({\theta}_t)$ means the ability estimate ${\theta}_t$ is more accurate and closer to real ability.


Intuitively, a stimulation should be given if the selected question by the selection algorithm helps enhance the accuracy of ability estimate. Comparatively, a punishment should be given if the selected question reduces this accuracy. Formally, we design the quality reward as:
\begin{equation}
    r_{qua} = ACC({\theta}_t) - ACC({\theta}_{t-1})
\label{equ:qualityreward}
\end{equation}
The meta question set is used to compute quality reward and will not be selected to the student. 

\subsubsection{\textbf{Diversity}}
\label{sec:diveristyreward}

In the large and comprehensive exam, the test questions should include rich knowledge concepts \cite{maat}. The diversity objective requires to cover a variety of knowledge concepts. This implies that a stimulation should be given if the selection algorithm chooses a question with a new concept. Here, for simplicity, we discretize the reward value. Formally, the diversity reward is 1 if a question involving a new concept is selected, and 0 otherwise:
\begin{equation}
    r_{div} =
    	\begin{cases}
				 1, \quad \text{if}\quad c_t \setminus \{c_1 \cup c_2 \ldots \cup c_{t-1}\} \neq \emptyset \\
				 0, \quad \text{otherwise}
		\end{cases}
		\label{equ:diversityreward}
\end{equation}
where $c_t$ is the concept covered by the question $q_t$ at test step $t$. The binary reward setting is commonly used in RL \cite{mooinlearning, morlrs}.

\subsubsection{\textbf{Novelty}}
As mentioned before, a proper selection algorithm should take novelty into account because the lack of novelty can lead to overexposed questions and may affect students' test behaviors \cite{han2018components}. For example, students can tell the later classmates the answers to the overexposed questions. 

Therefore, we design the novelty reward to control the question exposure. Let $\mathcal{T}$ denote the set of top $x\%$ of most popular questions in the training set. We encourage the selection of less popular questions, as these questions are more likely to be novel in the future and lead to a balanced distribution of question exposures. Along this line, the novelty reward is 1 if the selected question $q_t$ is not in $\mathcal{T}$, and 0 otherwise: 
\vspace{-5pt}
\begin{equation}
    r_{nov} =
    	\begin{cases}
				 1, \quad \text{if}\ q_t \notin 
 \mathcal{T}\\
				 0, \quad \text{otherwise}
		\end{cases}
		\label{equ:noveltyreward}
\end{equation}
Take note that after determining the training set, $\mathcal{T}$ is calculated and fixed, and will not change throughout the CAT process. In this paper, we set $x=10$ following \citet{morlrs}. 

So far, we have defined the vector-valued reward $\mathbf{r}(\mathbf{s}_t, q_t)= [r_{qua}, r_{div},r_{nov}]$, which refers to the quality, diversity and novelty reward respectively. From the optimization function in Equation \ref{eq:objectivefunction}, the reward vector is multiplied by the scalarization function $\textbf{w}$. The each element of $\textbf{w}$ can be regarded as the importance of the corresponding objective, and we discuss its role in the Section \ref{sec:objectivecomparion}.

\subsection{State Encoder}
\label{sec:stateencoder}
Generally, state encoder takes the historical response records as input to generate the state of GMOCAT: $\mathbf{s}_t=f_{se}(\{(q_1,c_1,y_1),\ldots,\\(q_{t-1},c_{t-1},y_{t-1})\})$. We will detail how the state encoder $f_{se}$ is implemented below.


Suppose the number of questions is $Q$, we use embedding matrix $\mathbf{W}_q \in \mathbb{R}^{Q\times d}$ to map each question $q$ into the real-valued embedding $\mathcal{E}_q \in \mathbb{R}^d$: $\mathcal{E}_q = \mathbf{x}_q \mathbf{W}_q$, where $\mathbf{x}_q$ is the one-hot vector of $q$, and $d$ is the embedding dimension. $\mathcal{E}_q$ characterizes information about questions. We apply the same operation to get the concept embedding $\mathcal{E}_c \in \mathbb{R}^d$ of concept $c$ and response embedding $\mathcal{E}_y \in \mathbb{R}^d$ of response $y$.

The relation aggregator is used to extract relational information and learn relation-aware embeddings. By using graph neural networks, the aggregator takes raw question embeddings $\mathcal{E}_q$ and raw concept embeddings $\mathcal{E}_c$ as input and outputs corresponding relation-aware embeddings $\widetilde{\mathcal{E}}_q$ and $\widetilde{\mathcal{E}}_c$ (detailed in the Section \ref{sec:relationaggregator}). Alternatively, we could use raw embeddings of questions and concepts similar to prior work, but we find that the relation-aware representations perform better in most datasets. The performance of relation-aware representations will be further discussed in Section \ref{sec:ablation}.



Given the step $t$, the state encoder represents historical question sequence $\{q_{1:t-1}\}$ by relation-aware question embeddings\footnote{For questions, $\{q_{1:t-1}\}$ is short for $\{q_1,\ldots,q_{t-1}\}$, and $\{\widetilde{\mathcal{E}}_q^{1:t-1}\}$ is short for $\{\widetilde{\mathcal{E}}_q^{1},\ldots,\widetilde{\mathcal{E}}_q^{t-1}\}$. This also applies to concepts and responses.} $\{\widetilde{\mathcal{E}}_q^{1:t-1}\}$, represents historical concept sequence $\{c_{1:t-1}\}$ by relation-aware concept embeddings $\{\widetilde{\mathcal{E}}_c^{1:t-1}\}$. The historical response sequence $\{y_{1:t-1}\}$ is represented by raw response embeddings $\{\mathcal{E}_y^{1:t-1}\}$. Note that the state encoder's input $\{(q_{t'},c_{t'},y_{t'})|t'\in [1,t-1]\}$ is the response records at all previous steps. Thus, for each previous step $t'\in [1,t-1]$, the triple $(q_{t'},c_{t'},y_{t'})$ can be represented by $\mathbf{e}_{t'}$, which is the concatenation of relation-aware question embedding $\widetilde{\mathcal{E}}_q^{t'}$, relation-aware concept embedding $\widetilde{\mathcal{E}}_c^{t'}$ and response embedding $\mathcal{E}_y^{t'}$. The formulation of $\mathbf{e}_{t'} \in \mathbb{R}^{D}$ is given by:
\begin{equation}
    \mathbf{e}_{t'} = \widetilde{\mathcal{E}}_q^{t'} \oplus \widetilde{\mathcal{E}}_c^{t'} \oplus \mathcal{E}_y^{t'}
\end{equation}
where $D=3d$. If a question contains multiple concepts, we take the mean of relation-aware embeddings of related concepts as $\widetilde{\mathcal{E}}_c^{t'}$.

Then the historical response records $\{(q_{t'},c_{t'},y_{t'})|t'\in [1,t-1]\}$ can be represented by a embedding matrix $\mathbf{E}_t =[\mathbf{e}_1, \mathbf{e}_2, \ldots, \mathbf{e}_{t-1}]^T \in \mathbb{R}^{(t-1)\times D}$. We note that response records contain different amounts of information. For example, answering a hard question correctly contains more information than answering a simple question correctly. To capture the differences among response records, we apply self-attention mechanism \cite{mhsa} on  $\mathbf{E}_t$, which is defined as the scaled dot-product function:

\begin{equation}
    \widetilde{\mathbf{E}}_t=
    \text{Softmax}(\frac{(\mathbf{E}_t \mathbf{W}^Q) (\mathbf{E}_t \mathbf{W}^K)^T}{\sqrt{d_k}})(\mathbf{E}_t \mathbf{W}^V)
    \label{eq:selfattention}
\end{equation}
where $\mathbf{W}^Q, \mathbf{W}^K, \mathbf{W}^V \in \mathbb{R}^{D \times D}$ are trainable matrices, $\sqrt{d_k}$ is scaling factor \cite{mhsa}. We add LayerNorm \cite{layernorm} and skip-connection \cite{skipconnect} behind the self-attention mechanism. We also use Dropout \cite{dropout} to avoid the overfitting.

Notice that the student’s real ability is unchanged during the CAT process \cite{cat:primer}, the order of each record is not important. We input $\widetilde{\mathbf{E}}_t$ (the embeddings after self-attention) into the average-pooling, and generate the state $\mathbf{s}_t\in\mathbf{R}^D$. The actor-critic recommender uses this state to select the next question.

\subsection{Actor-Critic Recommender}
\label{moppo}

The actor-critic recommender takes the state $\mathbf{s}_t$ as input. We utilize a policy network as the actor to generate actions by sampling from the distribution $\pi(q_t|\mathbf{s}_t;\phi_\pi)$. The actor is a fully connected layer with parameter $\phi_\pi$. Besides, we use a value network as the critic to evaluate the state. Given a state $\mathbf{s}_t$, the critic's output $\mathbf{V}(\mathbf{s}_t)$ is a vector that predicts the expected return, each element of which corresponds to an objective, defined by $\mathbf{V}(\mathbf{s}_t;\phi_v) = [V(\mathbf{s}_t)_{qua}, V(\mathbf{s}_t)_{div}, V(\mathbf{s}_t)_{nov}]$. The subscripts $\{qua, div, nov\}$ refer to quality, diversity and novelty objectives, respectively. The critic is also a fully connected layer with parameter $\phi_v$.

To maximize the weighted-sum return $\mathcal{J}$ in Equation \ref{eq:objectivefunction}, we modify PPO method \cite{ppo} into a multi-objective form. Specifically, the advantage value for selecting $q_t$ is defined as the actual return of a state-action pair minus the expected return of this state:
\begin{equation}
    \mathbf{A}(\mathbf{s}_t,q_t)=\sum_{t'=t}\gamma^{t'-t}\mathbf{r}(\mathbf{s}_{t'},q_{t'})-\mathbf{V}(\mathbf{s}_t)
    \label{eq:return}
\end{equation}
\vspace{-5pt}

We use the scalarization function $f_w$ to convert the vectorized advantage $\mathbf{A}$ into a scalar. As introduced before, we choose linear function $f_w=\mathbf{w}$, where importance weight $w_i$ corresponds to individual objective. The clipped surrogate loss is applied to update the actor parameters:
\vspace{-5pt}
\begin{equation}
\begin{split}
\mathcal{L}_1 = -& \mathbb{E}_{\tau \sim \pi_{old}} [\text{Min} \{\frac{\pi(q_t|\mathbf{s}_t)}{\pi_{old}(q_t|\mathbf{s}_t)} \mathbf{w}^T \mathbf{A}(\mathbf{s}_t,q_t),  \\ 
&\text{Clip} \left( \frac{\pi(q_t|\mathbf{s}_t)}{\pi_{old}(q_t|\mathbf{s}_t)},1-\epsilon , 1+\epsilon \right) \mathbf{w}^T \mathbf{A}(\mathbf{s}_t,q_t) \}] 
\end{split}
\label{eq:actorloss}
\end{equation}


The critic loss is based on the purpose that the expected return gets as close to the actual return as possible:
\begin{equation}
    \mathcal{L}_2=\frac{1}{2} \mathbf{w}^T \|\mathbf{V}(\mathbf{s}_t) - \sum_{t'=t}\gamma^{t'-t}\mathbf{r}(\mathbf{s}_{t'},q_{t'})\|^2 
\end{equation}

Finally, the Multi-Objective PPO(MOPPO) loss is a weighted sum of two losses with the trade-off hyperparameter $\alpha \in\mathbb R_{+}$:
\begin{equation}
    \mathcal{L} = \mathcal{L}_1 + \alpha \mathcal{L}_2
    \label{equ:moppoloss}
\end{equation}

\subsection{Relation Aggregator}
\label{sec:relationaggregator}

As we discussed above, the relational information included in the questions and concepts is closely related to our three objectives, but is neglected by existing CAT methods. We use the relation aggregator to aggregate relational information. In this part, we describe the aggregation of relational information from the perspectives of concepts and questions. 

\subsubsection{\textbf{Concept Relation}}
Since the concepts appear in both prerequisite and correlation graphs, concept embedding is influenced by two relations. We apply graph attention network \cite{gat} (GAT) to aggregate neighbor embeddings in two graphs. For each concept $c$ with raw embedding $\mathcal{E}_c$, let $N_c^{pre}, N_c^{cor}$ be its neighborhood in the prerequisite and correlation graph. We aggregate neighbor embeddings with the attention weights to get the prerequisite-aware embedding $\mathbf{g}_{pre}$ and the correlation-aware embedding $\mathbf{g}_{cor}$:
\begin{equation}
    \mathbf{g}_{pre} =\sum_{c'\in N_c^{pre}} \alpha_{c,c'}\mathbf{W}_{pre}\mathcal{E}_{c'},\ \  \mathbf{g}_{cor} =\sum_{q'\in N_c^{cor}} \beta_{c,q'}\mathbf{W}_{cor}\mathcal{E}_{q'}
\end{equation}
Intuitively, the attention weights ($\alpha_{c,c'}$ or $\beta_{c,q'}$) are related to the similarity between neighbor embeddings and the concept $c$ embedding, defined by:
\begin{align}
    \alpha_{c,c'} &= \text{Softmax}_{c'} \left( \text{att}_{pre} \left([\mathbf{W}_{pre}\mathcal{E}_{c},  \mathbf{W}_{pre}\mathcal{E}_{c'}]\right)\right), c' \in N_c^{pre} \\
    \beta_{c,q'} &= \text{Softmax}_{q'} \left( \text{att}_{cor} \left([\mathbf{W}_{cor}\mathcal{E}_{c}, \mathbf{W}_{cor}\mathcal{E}_{q'}]\right)\right), q' \in N_c^{cor} \label{eq:betaweight}
\end{align}
where $\text{att}_\bullet$ denotes a linear layer with a LeakyReLU activation function. $[\cdot]$ is the concatenation operation, $\mathbf{W}_{pre}, \mathbf{W}_{cor}$ are trainable matrices.

Prerequisite-aware embedding and correlation-aware embedding contain different relational information. To distinguish their different importances, we use the following treatment. Prerequisite-aware embedding's weight $\mu_{pre}$ is related to the similarity between the attention vector $\mathbf{P}$ and prerequisite-aware embedding, defined by 
\begin{equation}
    \mu_{pre} = \mathbf{P}^T \cdot tanh(\mathbf{W} \cdot \mathbf{g}_{pre}+\mathbf{b})
\end{equation}
and the weight $\mu_{cor}$ can be modeled similarly. These two weights are normalized by a softmax operation. Finally, the relation-aware embedding of concept $c$, denoted as $\widetilde{\mathcal{E}}_{c}$, is the weighted sum of prerequisite-aware embedding and correlation-aware embedding:
\begin{equation}
    \widetilde{\mathcal{E}}_{c}= \mu_{pre} \mathbf{g}_{pre} + \mu_{cor} \mathbf{g}_{cor} 
\end{equation}

\subsubsection{\textbf{Question Relation}}
Similar to the concept, we perform question neighbor aggregation via GAT again. For each question $q$ with raw embedding $\mathcal{E}_q$, let $N_q^{cor}$ be its neighbor set in the correlation graph. Since questions only contain correlation relation, the relation-aware embedding $\widetilde{\mathcal{E}}_{q}$ is the correlation-aware embedding $\mathbf{h}_{cor}$. The relation-aware embedding of question $q$ is $\widetilde{\mathcal{E}}_{q}$, given as:
\begin{gather}
    \mathbf{h}_{cor}=\sum_{c'\in N_q^{cor}} \xi_{q,c'}{\mathbf{W}}_{cor}\mathcal{E}_{c'}, \quad \widetilde{\mathcal{E}}_{q} = \mathbf{h}_{cor}
\end{gather}
The weight $\xi_{q,c'}$ can be modeled similar to the forms in Eq. \ref{eq:betaweight}.

\section{Experiments}
In this section, we conduct extensive experiments on three real-world educational datasets to evaluate the effectiveness of our proposed GMOCAT method.

\subsection{Data Partition and Experiment Process}
We evaluate our GMOCAT method on three real-world educational datasets: Eedi\footnote{https://eedi.com/projects/neurips-education-challenge}, ASSIST\footnote{https://sites.google.com/site/assistmentsdata/home/assistment-2009-2010-data/skill-builder-data-2009-2010} and Junyi\footnote{https://www.kaggle.com/datasets/junyiacademy/learning-activity-public-dataset-by-junyi-academy}. Eedi \cite{eedidataset} is from the response logs over the 2018-2020 years on an educational platform Eedi. ASSIST \cite{assist09} is from the ASSISTments online tutoring platform. Junyi \cite{junyi} is gathered from the exercise logs over the 2018-2019 year on the online learning website Junyi Academy.

For all datasets, we remove students with fewer than 40 test records. The statistics of the processed datasets are listed in the Table \ref{tab:statistics}. We use 80\%-10\%-10\% students for training, validation and testing respectively. The students in the training set do not appear in the validation/testing set. The training set is used to get calibrated question parameters, most popular questions for novelty reward, and to optimize the selection algorithm \footnote{The static selection algorithms do not require training.}. Furthermore, we partition the samples which contain student $i$'s test records into the candidate question set ($\mathcal{D}^i_c$, 80\%) and the meta question set ($\mathcal{D}^i_m$, 20\%). These two sets are not the same for each student, and also generated randomly in each training epoch to prevent overfitting \cite{bobcat}.

The experimental results are averaged over five runs. All our experimental results are obtained on the testing set. For each student $j$ in the testing set, (1) we use the selection algorithm to select a question from $\mathcal{D}^j_c$; (2) CDM updates ability estimate with the corresponding responses; (3) we evaluate the selection algorithm by different metrics and report the value. 


\begin{table}[]
    \centering
    \caption{Dataset Statistics}
    \vspace{-3mm}
    \begin{tabular}{l|c|c|c}
    \toprule
        Dataset & Eedi & ASSIST & Junyi \\ \hline
        \#Students & 4,918 & 1,360 & 20,395  \\ \hline
        \#Questions & 948 & 17,751 & 2,835\\ \hline
        \#Concepts & 86 & 123 & 40 \\ \hline
        \#Records & 1,382,727 & 239,919 & 2,537,898\\ \hline
        \#Prerequisite Edges & 334 & 1,166 & 306 \\ \hline
        Concepts Per Question & 4.0 & 1.2 & 1.0 \\ \hline
        Positive Label Rate & 0.55 & 0.62 & 0.69 \\
    \bottomrule
    \end{tabular}
    \vspace{-3mm}
    \label{tab:statistics}
\end{table}

\subsection{Evaluation Method}
\subsubsection{\textbf{Quality Metric}}

We evaluate the accuracy of final ability estimate of student $i$ by predicting binary-valued student responses on her meta question set $\mathcal{D}^i_m$. Therefore, we take Area Under ROC Curve (AUC) \cite{auc} and Accuracy (ACC) as quality metrics to assess different selection algorithms.


\subsubsection{\textbf{Diversity Metric}}
We use the concept coverage ($Cov$) \cite{maat} to measure diversity. Specifically, let $\mathcal{K}$ be concept set, and $\mathcal{K}_t$ be the set of concepts covered by all selected questions before step $t$. $Cov$ is defined as the proportion of covered concepts by all selected questions:
\vspace{-5pt}
\begin{equation} \label{eq:div-cov}
    Cov = \frac{1}{|K|}\sum_{k\in K} \mathbbm{1}(k \in \mathcal{K}_t)
\end{equation}

\subsubsection{\textbf{Novelty Metric}}
We measure the novelty using the question exposure rate (the proportion of times a question was selected) and the mean overlap rate (the mean overlap among questions selected by any two students in the student set) \cite{expandoverlap}:
\begin{align}
    &\text{Exposure}_q = \frac{N_q}{|\mathcal{U}|} \\
    \text{Overlap} &= \frac{\sum\sum_{i,j \in \mathcal{U},j\neq i} |Q_i \bigcap Q_j |}{ |\mathcal{U}|*(|\mathcal{U}|-1)/2  }
\end{align}
where $N_q$ is the count that question $q$ is chosen, $\mathcal{U}$ is the student set, $Q_i$ is the set of questions tested by student $i$.


\subsection{Baselines}
We apply our method on two CDMs: traditional Item Response Theory (IRT) \cite{irt} and recent deep learning-based model NeuralCDM (NCD) \cite{ncd}, and we compare our methods with two groups of selection algorithms. We use the following state-of-the-art selection algorithms as baselines. They are:

\textit{Static methods}. based on heuristic and unlearnable rules.
\begin{itemize}[leftmargin=2pt]
    \item Random: the random selection algorithm.
    \item MFI \cite{appofirt}: It selects the question with Maximum Fisher Information. It's only designed for IRT.
    \item KLI \cite{kli}: It selects the question with the maximum moving average of Kullback-Leibler information. It's only designed for IRT.
    \item MAAT \cite{maat}: It's an active learning-based method, which uses Expected Model Change (EMC) of CDM to select questions. It also designs an extra module to enhance concept diversity.
\end{itemize}

\textit{Learnable methods}. They are data-driven and learnable from large-scale datasets.
\begin{itemize}[leftmargin=2pt]
    \item BOBCAT \cite{bobcat}: It's a meta learning-based method that recasts CAT as a bilevel optimization problem.
    \item NCAT \cite{ncat}: It's a reinforcement learning-based method that designs an attention-based DQN. It selects questions by sampling from the Boltzmann distribution of Q values to control question exposure.
\end{itemize}

\begin{table*}[h]
\caption{The AUC and ACC performance on three public datasets. The best performance is in bold, while the second best value
is underlined. "-" indicates the method can't be applied on NCD. "$\ast$" indicates statistically significant improvement (measured by t-test) with p-value < 0.05.} 
\vspace{-5pt}
\resizebox{\linewidth}{!}{
\begin{tabular}{ll|cccccc|cccccc|cccccc}
\toprule
\multicolumn{2}{l|}{Dataset}   & \multicolumn{6}{c|}{\textbf{Eedi}}& \multicolumn{6}{c|}{\textbf{ASSIST}} & \multicolumn{6}{c}{\textbf{Junyi}}  \\ \hline
\multicolumn{2}{l|}{CDM}   & \multicolumn{3}{c}{IRT}    & \multicolumn{3}{c|}{NCD} & \multicolumn{3}{c}{IRT}    & \multicolumn{3}{c|}{NCD} & \multicolumn{3}{c}{IRT}    & \multicolumn{3}{c}{NCD}                  \\ \hline
\multicolumn{2}{l|}{Metric}& \multicolumn{6}{c|}{AUC}& \multicolumn{6}{c|}{AUC}& \multicolumn{6}{c}{AUC}   \\ \hline
\multicolumn{2}{l|}{Step}  & 5  & 10 & \multicolumn{1}{c|}{20} & 5  & 10 & 20  & 5  & 10 & \multicolumn{1}{c|}{20} & 5  & 10 & 20  & 5  & 10 & \multicolumn{1}{c|}{20} & 5  & 10 & 20   \\ \hline 
\multicolumn{1}{l|}{\multirow{2}{*}[-12pt]{Static}} & Random & 68.38&	69.73&	 \multicolumn{1}{c|}{71.98}  & 68.45 &	70.24 &	72.98   & 67.68 & 67.89 & \multicolumn{1}{c|}{ 68.43 }  & 67.73 & 68.51 & 69.70 & 76.72&	76.99&	\multicolumn{1}{c|}{77.44}   &76.80 &	77.06&	77.47 \\ \cline{2-20} 
\multicolumn{1}{l|}{} & MFI &  68.92 &	70.41 &	 \multicolumn{1}{l|}{72.66}   & - & - & - & 67.95 & 68.42 & \multicolumn{1}{l|}{69.26}   & - & - & -  & 77.03&	77.48&	 \multicolumn{1}{l|}{78.16}   & - & - & - \\ \cline{2-20} 
\multicolumn{1}{l|}{} & KLI &  68.69 &	70.29 &	  \multicolumn{1}{l|}{72.60}   & - & - & - &  67.92 & 68.39 &  \multicolumn{1}{l|}{69.23}   & - & - & - & 76.98&	77.43&	\multicolumn{1}{l|}{78.14}   & - & - & - \\\cline{2-20} 
\multicolumn{1}{l|}{} & MAAT &  \underline{69.09} &	\underline{70.90} &	  \multicolumn{1}{l|}{73.19}   & 69.03 &	71.03 &  73.75 & 68.24 & 68.82 &  \multicolumn{1}{l|}{69.70}   & 67.96 & 69.38 &  71.17 & 76.93&	77.39&	 \multicolumn{1}{l|}{78.21}   & 77.07&	77.53&	 78.40\\ \hline
\multicolumn{1}{l|}{\multirow{2}{*}[-6pt]{Learnable}}& BOBCAT & 68.94 &	70.50 & \multicolumn{1}{l|}{73.24}   &  \underline{69.17} &	71.44 &	 74.51 & 68.65 & 69.44 &  \multicolumn{1}{l|}{70.97}   & \underline{69.50} & 70.63 & \underline{71.80} & 77.81&	78.70 &	 \multicolumn{1}{l|}{\underline{79.17}}   & 77.47&	78.21&	79.46 \\  \cline{2-20}
\multicolumn{1}{l|}{} & NCAT   & 69.04 &	70.78 &	 \multicolumn{1}{l|}{\underline{73.32}}   & 69.09 &	\underline{71.45} &	\underline{74.55}   & \underline{68.67} & \underline{69.49} &  \multicolumn{1}{l|}{\underline{71.06}}   & 69.28 & \underline{70.93} &	71.68  & \underline{77.96} & \underline{78.87} &	 \multicolumn{1}{c|}{79.13}   & \underline{77.63} &	\underline{78.41} &	\underline{79.56}\\ \cline{2-20}
\multicolumn{1}{l|}{} & GMOCAT   & \textbf{69.81*} &	\textbf{71.78*} &	  \multicolumn{1}{l|}{\textbf{74.19*}}   & \textbf{71.25*} &	\textbf{73.76*} &	\textbf{75.76*}  & \textbf{69.13*} & \textbf{70.38*} &  \multicolumn{1}{l|}{\textbf{71.91*}}  & \textbf{69.95*} & \textbf{71.26*} & \textbf{72.95*} & \textbf{78.68*} &	\textbf{80.00*} &	 \multicolumn{1}{c|}{\textbf{80.24*}}   & \textbf{78.07*} &	\textbf{79.00*} &	\textbf{80.30*} \\ \hline \hline
\multicolumn{2}{l|}{Metric}& \multicolumn{6}{c|}{ACC}& \multicolumn{6}{c|}{ACC}& \multicolumn{6}{c}{ACC}   \\ \hline
\multicolumn{2}{l|}{Step}  & 5  & 10 & \multicolumn{1}{c|}{20} & 5  & 10 & 20  & 5  & 10 & \multicolumn{1}{c|}{20} & 5  & 10 & 20  & 5  & 10 & \multicolumn{1}{c|}{20} & 5  & 10 & 20   \\ \hline 
\multicolumn{1}{l|}{\multirow{2}{*}[-12pt]{Static}} & Random & 63.52 &	64.36 &	 \multicolumn{1}{c|}{65.91}  & 63.45	& 64.83 &	66.70 & 64.50 & 64.75 &  \multicolumn{1}{c|}{ 65.21 }  & 65.71&	66.29&	67.15 &72.66&	73.07&	 \multicolumn{1}{c|}{73.67}   &73.81&	74.00 &	74.36 \\ \cline{2-20} 
\multicolumn{1}{l|}{} & MFI &  63.79 &	64.63 &	 \multicolumn{1}{l|}{65.92}   & - & - & - & 64.83&	65.34&	 \multicolumn{1}{l|}{66.19}   & - & - & -  & 72.88&	73.34&	 \multicolumn{1}{l|}{73.98}   & - & - & - \\ \cline{2-20} 
\multicolumn{1}{l|}{} & KLI &  63.54 &	64.44 &	 \multicolumn{1}{l|}{65.85}   & - & - & - & 64.75&	65.29&	 \multicolumn{1}{l|}{66.16}   & - & - & - & 72.87&	73.33&	 \multicolumn{1}{l|}{73.97}   & - & - & - \\\cline{2-20} 
\multicolumn{1}{l|}{} & MAAT &  63.25 &	64.38 &	 \multicolumn{1}{l|}{66.15}   & 63.86 &	64.58 &	66.71 &  \underline{65.56} &	66.30 &	 \multicolumn{1}{l|}{67.57}   & 66.57&	67.84&	 \multicolumn{1}{l|}{69.52} & 73.73&	74.51&	 \multicolumn{1}{l|}{75.48}   & 74.31&	74.88&	 75.38 \\ \hline
\multicolumn{1}{l|}{\multirow{2}{*}[-6pt]{Learnable}}& BOBCAT & 63.91 & 64.90 &	 \multicolumn{1}{l|}{\underline{66.97}}   & 63.95 &	\underline{65.62} &	 67.69 & 65.37&	66.21&	  \multicolumn{1}{l|}{67.88}   & \underline{67.55} & \underline{68.63} & \underline{69.94} & 74.21&	75.66&	 \multicolumn{1}{l|}{\underline{76.51}}   & 74.42&	75.21&	\underline{76.45}\\  \cline{2-20}
\multicolumn{1}{l|}{} & NCAT   & \underline{64.04} & \underline{64.97} &	\multicolumn{1}{l|}{66.92}   & \underline{64.00} &	65.59 &	\underline{67.84}  & 65.34	& \underline{66.32} &	 \multicolumn{1}{l|}{\underline{68.36}}   & 67.15&	68.38&	69.44  & \underline{74.74} & \underline{76.05} & \multicolumn{1}{c|}{75.70}   &\underline{74.64} & \underline{75.47} &	76.40 \\ \cline{2-20}
\multicolumn{1}{l|}{} & GMOCAT   & \textbf{64.42*} &	\textbf{65.70*} &	\multicolumn{1}{l|}{\textbf{67.49*}}   & \textbf{65.53*} &	\textbf{67.36*} &	\textbf{68.96*} &  \textbf{66.16*}&	\textbf{67.42*}&	 \multicolumn{1}{l|}{\textbf{69.02*}}   & \textbf{67.63*}&	\textbf{68.55*}&	\textbf{70.18*} &\textbf{75.01*} &	\textbf{76.68*}&	\multicolumn{1}{c|}{\textbf{77.37*}}   &\textbf{74.84*}&	\textbf{75.83*}&	\textbf{77.16*} \\  \bottomrule 
\end{tabular}}

\label{tab:aucandacc}
\vspace{-5pt}
\end{table*}

\vspace{-5pt}
\subsection{Implementation Details}

For all experimental results, except for Section \ref{sec:objectivecomparion}, we always keep the scalarization function $\mathbf{w}=[1, 1, 1]$. The implementation of our model is available\footnote{The MindSpore implementation is available at: \url{https://gitee.com/mindspore/models/tree/master/research/recommend/GMOCAT}}. We set the maximum test length $T=20$, following \cite{ncat,rat}. We set $\gamma=0.5$ in Eq. \ref{eq:return}, $\epsilon=0.2$ in Eq. \ref{eq:actorloss}. The batch size is $128$, the embedding dimension $d=128$. The dropout rate is 0.1 in the self-attention mechanism. The optimizer is Adam \cite{adam}, and the learning rate is 0.001. The loss trade-off parameter $\alpha=1.0$. The parameters of the baselines all follow the settings in their original papers to ensure their best performance. In the three datasets, only Junyi provides the prerequisite relation between knowledge concepts. Therefore, we use the implementation by \citet{rcd} to construct the prerequisite graph for other two datasets, presented in Appendix \ref{app:graphconstruction}.

\vspace{-5pt}
\subsection{Performance Comparison}
Here we analyze the performance of GMOCAT on different metrics. The scalarization function $\mathbf{w}$ is fixed as [1, 1, 1].

\subsubsection{Quality Comparison}
Table \ref{tab:aucandacc} reports the AUC and ACC metrics of different methods at test step $t=5,10,20$. From them, we observe that: 
\begin{itemize}[leftmargin=12pt]
    \item[(1)] Our method outperforms all of the baselines on two different CDMs of three public datasets. On Eedi, GMOCAT achieves up to 2\% AUC improvements compared to the best baseline (e.g., step 5 on NCD). On ASSIST, GMOCAT achieves up to 1\% ACC improvements compared to the best baseline (e.g., step 10 on IRT). On Junyi, GMOCAT outperforms the best baseline by 1\% AUC points (e.g., step 20 on IRT). These results demonstrate that relational information and multi-objective strategies can improve the accuracy of ability estimates.
    \item[(2)] NCAT, also as a RL-based method, is the second best on most datasets, which also demonstrates the effectiveness of the RL framework. For example, at the beginning of test(step 5), MAAT is not weak, but at the end(step 20), its AUC/ACC are always beaten by the RL-based methods. This is because MAAT selects questions based on greed rather than a long-term perspective.
    
\end{itemize}

\subsubsection{Diversity Comparison}
\begin{figure}
    \subfigure[IRT on Eedi]{
    \centering
    \includegraphics[width=0.47\linewidth]{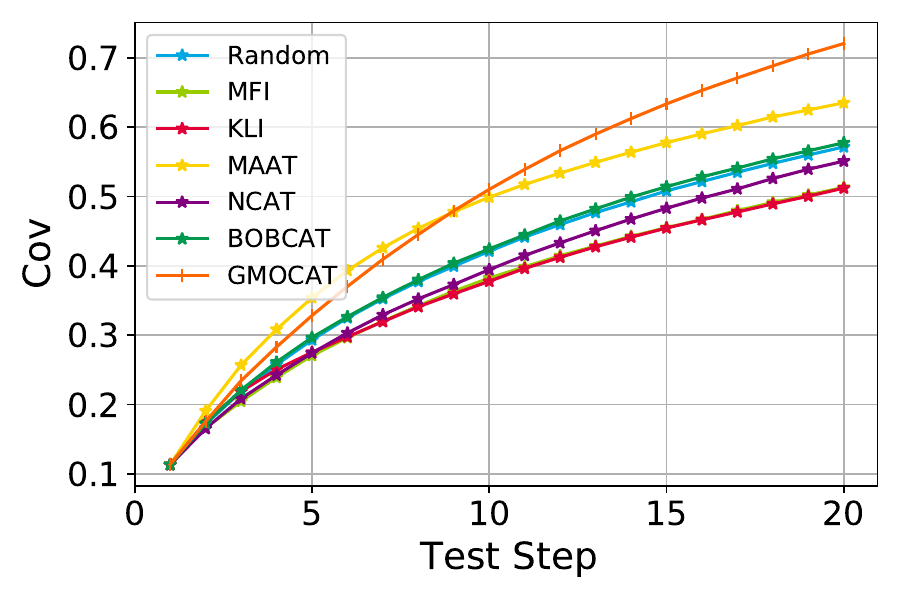}
    }
    \subfigure[NCD on Eedi]{
    \centering
    \includegraphics[width=0.47\linewidth]{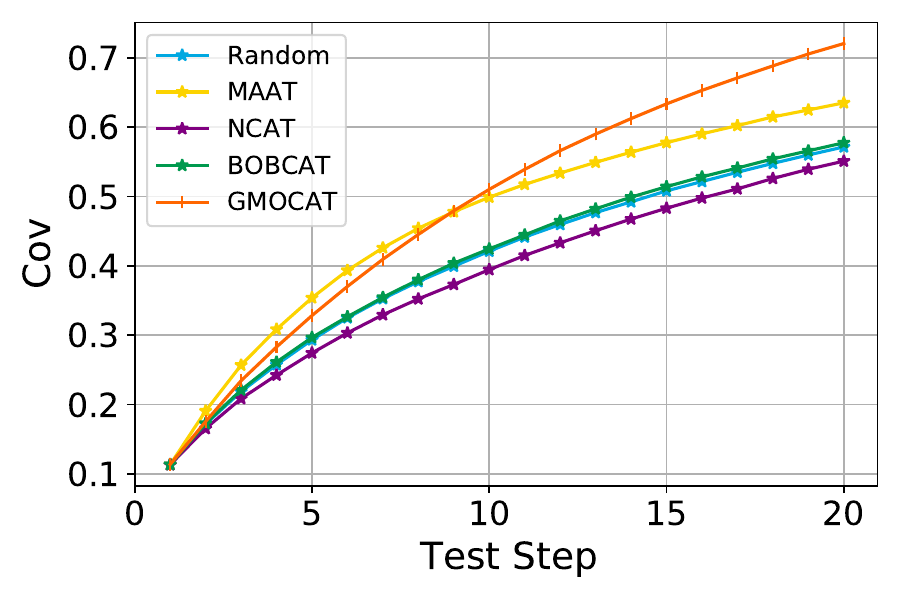}
    }
    
    \subfigure[IRT on ASSIST]{
    \centering
    \includegraphics[width=0.47\linewidth]{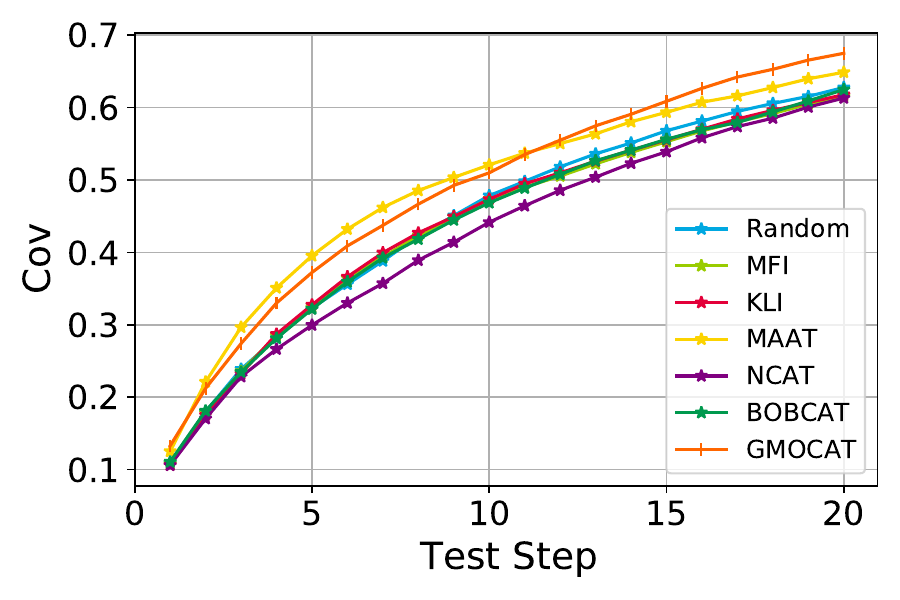}
    }
    \subfigure[NCD on ASSIST]{
    \centering
    \includegraphics[width=0.47\linewidth]{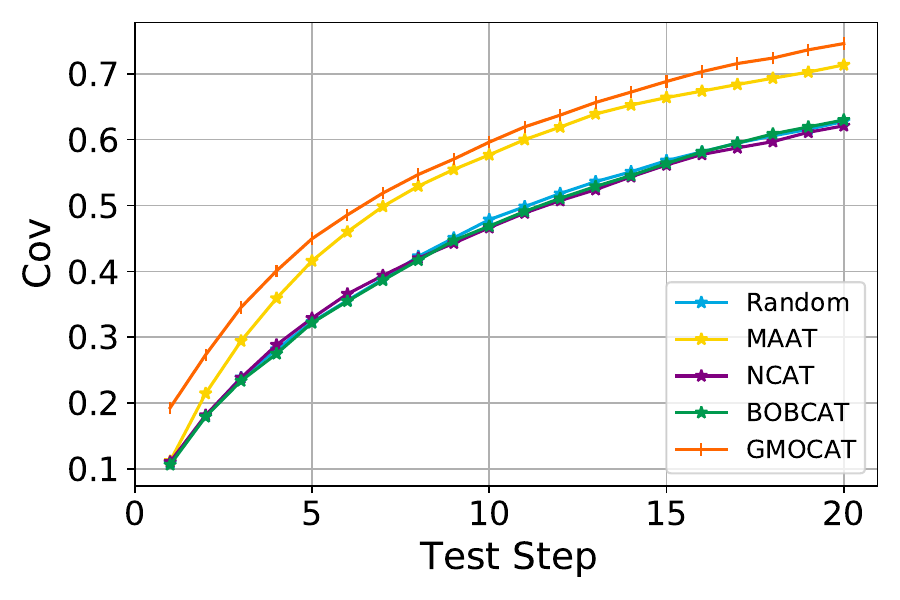}
    }
    
    \subfigure[IRT on Junyi]{
    \centering
    \includegraphics[width=0.47\linewidth]{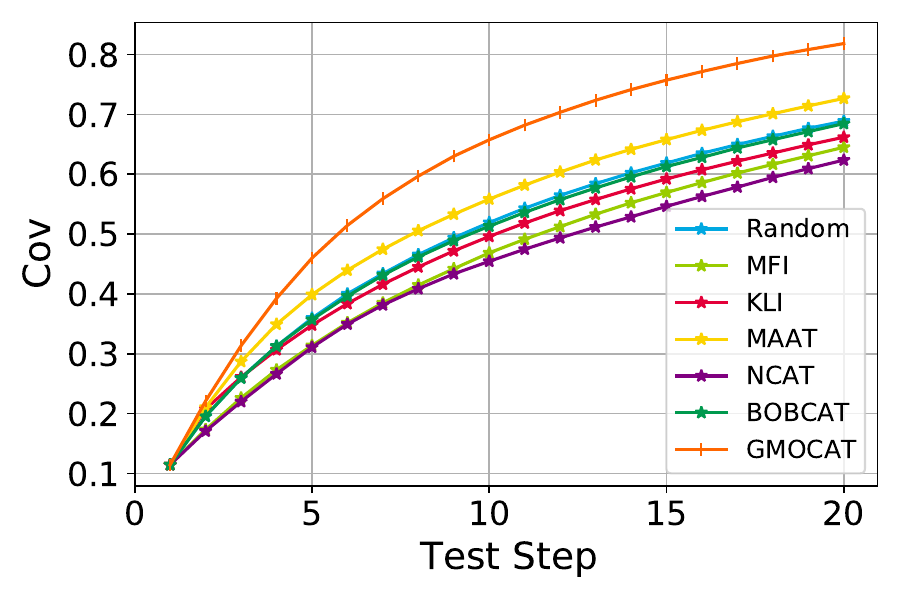}
    }
    \subfigure[NCD on Junyi]{
    \centering
    \includegraphics[width=0.47\linewidth]{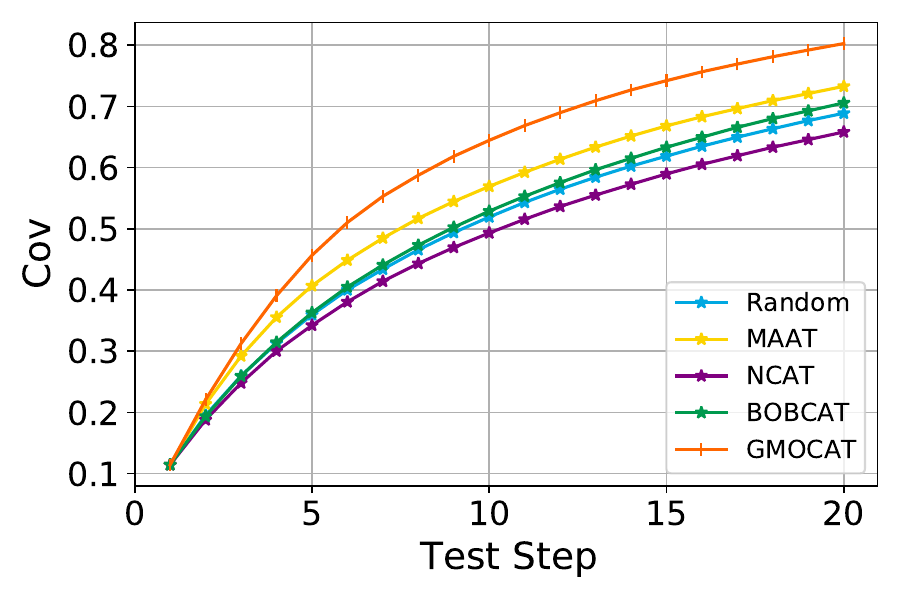}
    }
    \vspace{-7pt}
    \caption{Diversity Comparison with the Cov Metric.}
    \vspace{-15pt}
    \label{fig:cov}
\end{figure}

We display the $Cov$ curve throughout the CAT process in Figure \ref{fig:cov}. GMOCAT outperforms much on all datasets with two CDMs, because it has an explicit diversity objective in the MORL framework and a relation-aware selection algorithm, while other methods do not. Compared with other baselines, the Cov curve of GMOCAT grows fastest. In most cases, the second fastest method is MAAT, because it also has an intrinsic diversity goal. But its greedy strategy leads to performance degradation. 

\begin{figure*}
    \subfigure[(a) Quality: AUC@20]{
    \centering
    \includegraphics[width=0.25\linewidth]{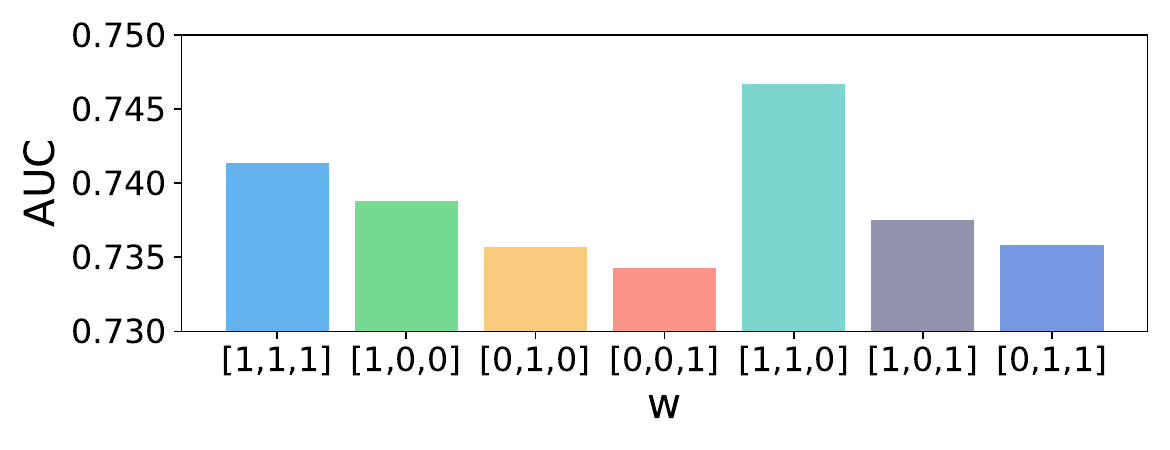}
    \label{fig:aucinw}
    }
    \hspace{-7pt}
    \subfigure[(b) Quality: ACC@20]{
    \centering
    \includegraphics[width=0.25\linewidth]{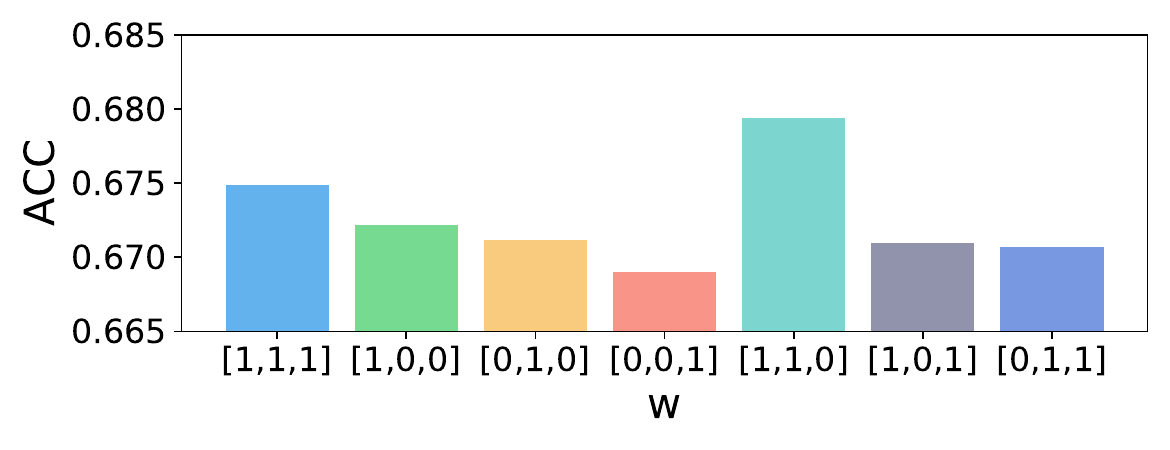}
    \label{fig:accinw}
    }
    \hspace{-7pt}
    \subfigure[(c) Diversity: Cov@20]{
    \centering
    \includegraphics[width=0.24\linewidth]{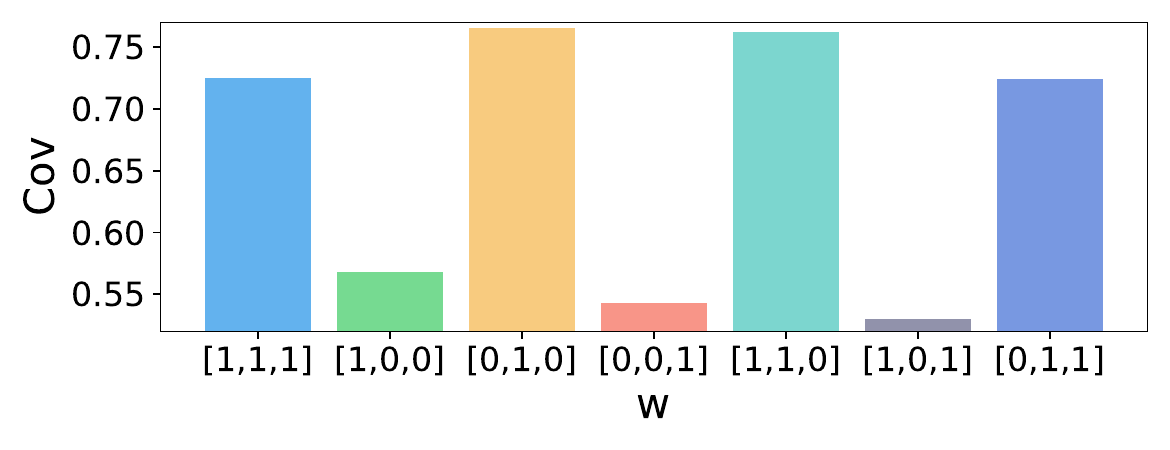}
    \label{fig:covinw}
    }
    \hspace{-7pt}
    \subfigure[(d) Novelty: Overlap@20]{
    \centering
    \includegraphics[width=0.24\linewidth]{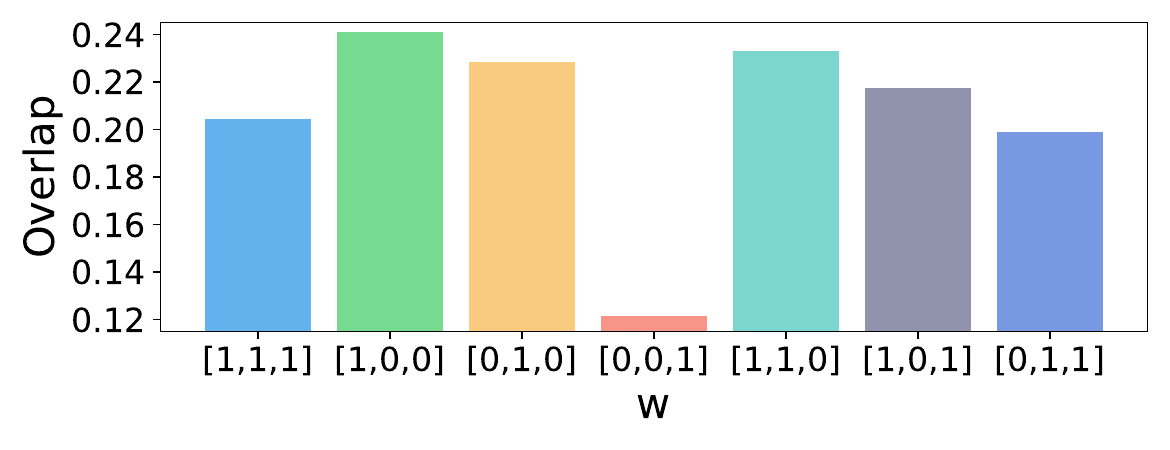}
    \label{fig:overinw}
    }
    \caption{Performance Comparison about GMOCAT using different configurations of $\mathbf{w}$, which cause to focus on a subset of objectives. The results are at step 20 on IRT with Eedi dataset. The first, second and third entries of $\mathbf{w}$ corresponds to the importance of quality, diversity and novelty objective, respectively.}
    \label{fig:rewardweight}
\end{figure*}
\subsubsection{Novelty Comparison}
Table \ref{tab:expandoverlap} lists the exposure rate(Exp.) and mean overlap rate(Over.) at step 20 with Junyi dataset, where Exp.(>0.2) represents the proportion of questions with exposure rate > 0.2. We find that: 
\begin{itemize}[leftmargin=12pt]
    \item[(1)] Although Random method has the lowest exposure rate, this does not mean that it is the best method. Because the randomly selected questions are not personalized at all, which violates the intention of CAT. We list Random method here mainly to illustrate the actual lower bound of the exposure rate.
    \item[(2)] Among all methods except Random, GMOCAT achieves the smallest exposure rate and overlap rate, which demonstrates the effectiveness of novelty reward. NCAT also gets a low question exposure because it samples actions from a Boltzmann distribution. In contrast, our approach achieves lower question exposure by directly targeting exposure as an optimization objective.
\end{itemize}

From the above three aspects, we can see that by using GMOCAT method, we can not only achieve a balance between accuracy, diversity and novelty, but also significantly improve all of quantitative metrics. Strengthening diversity and novelty objectives brings a notable improvement in $Cov$ and exposure rate metrics. 

\begin{table}[]

\caption{Exposure rate (Exp.) and mean overlap rate (Over.) at test step 20 with Junyi dataset. Exp.(>0.2) represents the proportion of questions with exposure rate > 0.2. The best result is in bold, while the second best is underlined. "$\ast$" indicates statistically significant improvement (measured by t-test) with p-value < 0.05.}
\vspace{-5pt}
\begin{threeparttable}
\resizebox{\linewidth}{!}{
\begin{tabular}{ll|cccc}
\toprule
\multicolumn{2}{l|}{Dataset}   & \multicolumn{4}{c}{\textbf{Junyi}}  \\ \hline
\multicolumn{2}{l|}{CDM}   & \multicolumn{2}{c}{IRT}  & \multicolumn{2}{c}{NCD}   \\ \hline
\multicolumn{2}{l|}{Metric}  & Exp.\%(>0.2) & \multicolumn{1}{c|}{Over.\%}  & Exp.\%(>0.2) & Over.\%  \\ \hline 
\multicolumn{2}{c|}{Random\tnote{a}}  &0.04 &	 \multicolumn{1}{c|}{4.6}  & 0.04 & 4.6 \\ \hline 
\multicolumn{1}{l|}{\multirow{2}{*}[-5pt]{Static}} & MFI &	\underline{0.18} &	 \multicolumn{1}{c|}{\underline{6.62}}  & - & -  \\ \cline{2-6}
\multicolumn{1}{l|}{} & KLI &	0.21 &	  \multicolumn{1}{c|}{6.73}   & - & -  \\\cline{2-6}
\multicolumn{1}{l|}{} & MAAT  &	0.63 &	  \multicolumn{1}{c|}{17.66}   &	0.67 &  15.22 \\ \hline
\multicolumn{1}{l|}{\multirow{2}{*}[-6pt]{Learnable}}& BOBCAT & 0.74 &  \multicolumn{1}{c|}{17.36}  &	0.88 &	17.64  \\  \cline{2-6}
\multicolumn{1}{l|}{} & NCAT  & 0.21 &  \multicolumn{1}{c|}{7.62}   &	\underline{0.14} &	 \underline{7.03}   \\ \cline{2-6}
\multicolumn{1}{l|}{} & GMOCAT  & \textbf{0.11*} & \multicolumn{1}{c|}{\textbf{5.13*}}   &	\textbf{0.07*} & \textbf{4.86*}  \\ \bottomrule
\end{tabular}}
\begin{tablenotes}
    \footnotesize
    \item[a] We list Random method here just to illustrate the actual lower bound of the\\ exposure rate.
  \end{tablenotes}
\end{threeparttable}
\label{tab:expandoverlap}
\vspace{-10pt}
\end{table}

\subsection{Ablation Study}
\label{sec:ablation}
We conduct ablation studies to further investigate the contribution of each module in GMOCAT. We test all metrics at step 20 on IRT with Junyi dataset. We still set the scalarization function $\mathbf{w} = [1,1,1]$. The settings are discussed as follows:
\begin{itemize}[leftmargin=8pt]
    \item GMOCAT-R: remove the relation-aware embedding. That means we ignore the relation graphs, and replace the relation-aware embedding $\widetilde{\mathcal{E}}_q,\widetilde{\mathcal{E}}_c$ with raw embedding $\mathcal{E}_q,\mathcal{E}_c$.
    \item GMOCAT-S: remove the scalarization function. That means the reward is changed from a vector to a scalar, weighted by three rewards. Correspondingly, the critic's output becomes a scalar.
\end{itemize}

The results are presented in Table \ref{tab:ablation}. We can observe that the GMOCAT's performance will decrease no matter which module is removed. This means that both modules contribute to the performance of GMOCAT. We analyze that: (1) GMOCAT-R loses vital relational information between questions and knowledge concepts, which significantly reduces its performance. This allows us to safely conclude that it is advisable to capture the relational information for selecting more appropriate questions. (2) after removing the scalarization function, the performance degradation of GMOCAT-S also proves the necessity of converting the single-objective method to the multi-objective method with vectorized rewards.

\begin{table}[]
\caption{The results of ablation studies. We test each metric at step 20 on IRT with Junyi dataset. The best results are given in bold. "$\ast$" indicates statistically significant improvement (measured by t-test) with p-value < 0.05.}
\begin{tabular}{l|ccc}
\toprule
Metric        &  AUC@20   &  Cov@20 & Overlap@20 \\ \hline
GMOCAT       &  \textbf{0.8024*} & \textbf{0.8185*} & \textbf{0.0513*} \\ \hline
GMOCAT-R     &  0.7999   &  0.7380    & 0.0599 \\ \hline
GMOCAT-S   & 0.7983 & 0.7310 & 0.0586 \\  \bottomrule
\end{tabular}
\vspace{-7pt}
\label{tab:ablation}
\end{table}

\subsection{Objective Comparison}
\label{sec:objectivecomparion}
To investigate the function of different objectives, in this part we explore the differences when GMOCAT focuses on the subsets of three objectives. In our setting, the first, second, and third entries of $\mathbf{w}$ correspond to the quality, diversity and novelty objectives respectively. We conduct the experiments with the following configurations of $\mathbf{w}$:
\begin{equation}
    \mathbf{w} \in \{[1,1,1], [1,0,0], [0,1,0], [0,0,1],[1,1,0],[1,0,1],[0,1,1] \}\nonumber
\end{equation}
Here, we mainly consider the effect of the presence/absence of each object, so the value of $\mathbf{w}$ is either 1 (presence) or 0 (absence).

Figure \ref{fig:rewardweight} displays the performance comparison of GMOCAT on Eedi+IRT setting at test step 20 with different configurations of $\mathbf{w}$. From the results we find some interesting phenomena:
\begin{itemize}[leftmargin=12pt]
    \item[(1)] Focusing solely on one objective leads to performance degradation in other metrics. For example, [1,0,0] gets a low coverage (Cov) value. [0,1,0] and [0,0,1] get low AUC/ACC values. This demonstrates the importance and necessity of utilizing multiple objectives simultaneously.
    \item[(2)] From Figure \ref{fig:aucinw} and \ref{fig:accinw}, we find that adding the diversity objective increases the value of AUC/ACC (e.g., [1,1,0] outperforms [1,0,0] on AUC/ACC). This phenomenon is consistent with our intuition, because the student's ability is multifaceted. We can predict the student's ability more accurately if test questions include diverse knowledge concepts.
    \item[(3)] From Figure \ref{fig:aucinw} and \ref{fig:accinw}, we also spot that adding novelty objective slightly impairs quality performance (e.g., [1,1,0] outperforms [1,1,1] on AUC/ACC). This phenomenon implies a potential conflict between quality and novelty objectives. The novelty objective attempts to achieve a balanced distribution of test questions, which results in low-quality questions being selected more frequently, and prevents us from predicting the student's ability accurately. 
\end{itemize}
From the above observation, we can conclude that there is both promotion and contradiction between three objectives. GMOCAT provides a flexible way to adapt CAT with different practical needs. For example, if we want to pay more attention to the exposure rate of test questions, we can increase the importance of novelty objectives while ensuring that quality performance does not degrade too much. We leave it to future work to quantitatively analyze the trade-off of multiple objectives and automate the balance to meet various practical applications.

\vspace{-5pt}
\section{Conclusion}
In this work, we propose GMOCAT, a Graph-Enhanced Multi-Objective method for CAT, that provides a multi-objective approach for learning the selection algorithm. We highlight the three objectives in CAT, namely quality, diversity and novelty, and apply Scalarized Multi-Objective RL to optimize these objectives in a long-term perspective. Furthermore, our method improves upon existing CAT methods by building relation-aware representations of questions and concepts to summarize their relational information. Extensive experiments demonstrated that our method outperforms state-of-the-art methods on three educational datasets, which significantly improves the accuracy of ability estimate, diversifies the test questions and alleviates the question exposure.


\begin{acks}
The SJTU team is supported by National Key R\&D Program of China (2022ZD0114804), Shanghai Municipal Science and Technology Major Project (2021SHZDZX0102) and National Natural Science Foundation of China (62076161, 62177033, U19A2065). The work is also sponsored by Huawei Innovation Research Program. We thank MindSpore \cite{mindspore} for the partial support of this work.
\end{acks}
\bibliographystyle{ACM-Reference-Format}
\bibliography{bibliography}

\newpage
\appendix
\section{The process of single-objective selection algorithm}
\label{app:training/testing_phase}
\begin{figure}[!h]
    \centering
    \includegraphics[width=\linewidth]{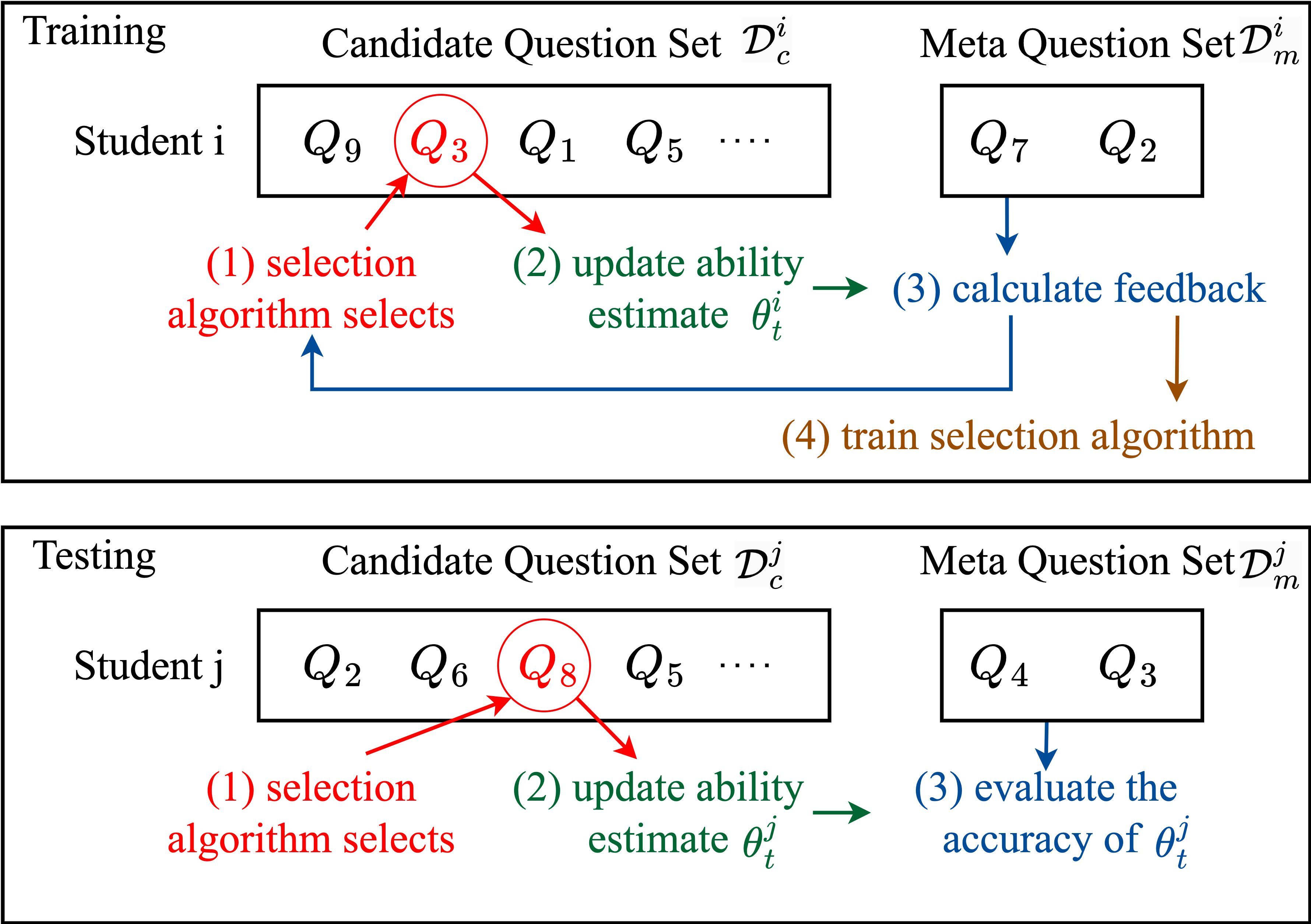}
    \caption{The training/testing of the single-objective selection algorithm.}
    \vspace{-10pt}
\end{figure}

\section{prerequisite Graph Construction}
\label{app:graphconstruction}
In the three datasets, only Junyi provides the prerequisite relation between knowledge concepts. Therefore, we need to construct the prerequisite graph for other two datasets. Here, we use the implementation by \citet{rcd}. First, we compute correct matrix $C$, where $C_{ij}$ is $\frac{n_{ij}}{\sum_k n_{ik}}$ if $i\neq j$ ,else it's 0. $n_{ij}$ means the count that concept $j$ is answered correctly and immediately after $i$ is answered correctly. Next, we calculate transition matrix $T$, where $T_{i,j}=1$ if $\frac{C_{ij}-min(C)}{max(C)-min(C)} > threshold$, representing concept $i$ has an edge pointing to $j$. If $T_{ij}=1$ but $T_{ji}\neq 1$, the relation between concept $i$ and $j$ is prerequisite, i.e., $i$ is a prerequisite for $j$. We set $threshold$ as the average value of matrix $T$.



\end{document}